\documentclass[aps,reprint,prd,superscriptaddress,nofootinbib,amsmath,amssymb]{revtex4-2}

\usepackage{amssymb}
\usepackage[utf8]{inputenc}
\usepackage{amsmath}
\usepackage{amsfonts}
\usepackage{color}
\usepackage{graphicx}
\usepackage{verbatim}
\usepackage{mathrsfs}
\usepackage{hyperref}

\begin{document}

\title{The lack of influence of the scalar hair on the DC conductivity.}

\author{Ulises \surname{Hernandez-Vera}}
\email[]{uhernandez.vera@gmail.com}
\affiliation{Instituto de Matem\'{a}ticas (INSTMAT), Universidad de Talca, Casilla 747, Talca 3460000, Chile}

\begin{abstract}
Recently obtained black hole solutions within the framework of beyond-Horndeski theories, which have the advantage of featuring primary hair, are generalized in the presence of two axionic fields. In order to induce a momentum dissipation, the axionic field solutions are homogeneously distributed along the horizon coordinates of the planar base manifold. We show that, despite the explicit dependence of the scalar field and the metric on the primary hair, this latter does not directly affect the calculation of transport properties. Its influence is indirect, modifying the horizon location, but the transport properties themselves do not explicitly depend on the hair parameter. We take a step further and show that even within a more general class of beyond-Horndeski theories, where the scalar field depends linearly on the hair parameter, the scalar hair still has no direct impact on the DC conductivity. This result underscores the robustness of our earlier findings, and seem to confirm that the transport properties remain unaffected by the explicit presence of the hair parameter.

\end{abstract}

\maketitle

\section{Introduction}

The study of black holes in alternative gravitational theories has gained significant importance, particularly in the exploration of solutions with "hair." While black holes are traditionally characterized by their mass, charge, and angular momentum, the concept of black hole hair introduces additional features which potentially modify the physical properties. Despite considerable theoretical advancements, the physical interpretation of hairy black hole solutions remains somewhat ambiguous in certain cases, particularly concerning the underlying nature of the hair parameter. On the other hand, understanding how the hair parameter can lead to observable consequences in the context of astrophysical black holes is undoubtedly a significant challenge.

In this context, we propose that the study of DC conductivity in holographic systems could provide valuable insights into the physical significance of (primary) hair. By examining how momentum dissipation and transport properties emerge in black hole spacetimes with (primary) hair, we aim to better understand the connection between the spacetime structure and the boundary field theory. Before going into the specifics of DC conductivity calculations we plan to do, we first provide a brief overview of the relationship between black holes and transport properties, particularly how these coefficients can be derived from the black hole horizon data. This will serve as a foundation for exploring the physical implications of primary hair of the axionic black hole solutions we will obtain.

In systems with perfect translational symmetry, the conservation of momentum prevents charge carriers from dissipating their momentum, leading to infinite DC conductivity. However, real-world materials exhibit finite conductivity because of mechanisms such as lattice vibrations or disorder, which break the translational invariance. These processes introduce momentum relaxation, resulting in a finite timescale over which momentum dissipates. In holography, momentum dissipation is modeled by breaking translational symmetry in the bulk spacetime, either explicitly through spatially dependent axion fields as we will do here, or periodically via Q-lattice scalar fields. Such models have proven invaluable for studying transport phenomena in strongly coupled systems. For instance, Q-lattice deformations or linear axions in holographic setups lead to finite DC conductivities in the dual field theory, mimicking the effects of resistivity in condensed matter systems; see \cite{Donos:2014cya, Donos:2013eha, Blake:2013bqa}. A significant advance in this field is the realization that DC transport coefficients such as electrical and thermal conductivities can be computed directly from the black hole horizon data. This approach, formalized in Refs. \cite{Donos:2014cya, Donos:2014cya, Andrade:2013gsa}, utilizes the near-horizon structure of the spacetime to express conductivities in terms of geometric and field quantities, bypassing the need to solve the full and complicated bulk equations. A convenient setup involves focusing on theories that admit axionic black hole solutions, which will be the approach we follow in this work. Examples of planar axionic black holes for scalar field nonminimally coupled together with two axionic field homogeneously distributed along the planar coordinates were shown to exist in \cite{Bardoux:2012aw, Bardoux:2012tr, Caldarelli:2013gqa}. On the other hand,  axionic black hole solutions are significant as gravitational duals for strongly coupled quantum field theories, particularly in systems where momentum dissipation will occur by choosing the axionic field homogeneously distributed along the horizon coordinates of the planar base manifold. In the current literature, there are numerous examples of axionic black hole solutions for which transport properties have been analyzed through this framework, particularly in the context of theories involving a scalar field. Without being exhaustive, we can mention cases such as a scalar field nonminimally (and possibly conformally) coupled \cite{Cisterna:2018hzf, Cisterna:2019uek}, for some Horndeski sectors \cite{Baggioli:2021ejg, Figueroa:2020tya}, for $k-$essence axionic terms \cite{Cisterna:2017jmv} or even in five dimensions with a Einstein-Gauss-Bonnet gravity theory \cite{Hao:2022zkr}.

To explore the influence that scalar hair could have from a holographic perspective, we propose as a first step to construct axionic black hole solutions that incorporate a scalar field explicitly dependent on the hair parameter. In these solutions, the hair parameter not only affects the scalar field but also appears explicitly in the metric itself, providing a direct connection between the geometry and the scalar dynamics, see Refs. \cite{Bakopoulos2024} and \cite{Baake:2023zsq, Bakopoulos:2023sdm} for such black hole solutions but without axionic fields. Axionic black holes with primary hair are particularly interesting because they can allow for a clear and testable framework to evaluate the role of scalar hair in holographic setups. These solutions arise naturally in certain scalar-tensor theories, particularly those known as beyond-Horndeski \cite{Gleyzes:2014dya, Gleyzes:2014qga} or DHOST theories \cite{Crisostomi:2016czh, BenAchour:2016fzp}. These extended models of gravity include additional terms in the Lagrangian, beyond the standard Horndeski framework \cite{Horndeski:1974wa}, which provide richer dynamics and allow for coupling mechanisms that make the hair parameter more prominent. In fact, unlike Horndeski theory, which guarantees second-order equations of motion, beyond-Horndeski or DHOST theories allow for higher-order equations. However, these theories are carefully constructed so that the propagating degrees of freedom remain well-defined, thanks to degeneracy conditions that eliminate ghost modes. For a review on black holes in scalar-tensor theories within the general framework of Horndeski theories, see \cite{Lecoeur:2024kwe, BenAchour:2024hbg}. By leveraging these extended theories, we aim to construct axionic black hole configurations where the presence of the scalar hair is encoded explicitly in the spacetime geometry as well as in the scalar field. Our ultimate goal is to use these solutions as a basis to analyze how the hair parameter might or not influence physical observables, such as the DC conductivity in holographic models.

The plan of the paper is organized as follows. In the next section, we will examine a model belonging to the beyond-Horndeski class, for which a black hole solution with primary hair was identified in \cite{Bakopoulos2024}, and latter generalized in \cite{Baake:2023zsq, Bakopoulos:2023sdm}. We extend these solutions by charging them through the Maxwell term and by introducing two axionic fields, enabling momentum relaxation. For this specific axionic solution, where the hair appears explicitly in both the metric and the scalar field, we will calculate the DC conductivity using the methods outlined in \cite{Donos:2013eha, Donos:2014cya}. We will demonstrate that the resulting expression for the DC conductivity does not explicitly depend on the hair parameter. The expectation of seeing the hair parameter appear in the expression for the DC conductivity was driven by the fact that the scalar field explicitly and linearly depended on the hair parameter as $\phi(t,r)=q t+\psi(r)$, where $q$ is the hair parameter. Building on this, we will consider in Section III a more general beyond-Horndeski theory that remains shift-invariant. This ensures that the scalar field can once again depend on the hair parameter as previously. However, as general as the functions of the beyond-Horndeski theory may be, we will observe that the DC conductivity remains insensitive to the hair parameter. Even in this case, we will show that it is impossible for the hair parameter to explicitly appear in the transport coefficients. The final section will be dedicated to our conclusions and future prospects.

\section{Charged axionic black holes with primary scalar hair}

As previously mentioned, we will focus on a simple scalar-tensor theory model that belongs to the class of so-called beyond-Horndeski theories. This model, along with certain extensions, has the notable feature of admitting spherically symmetric black hole solutions with primary hair \cite{Bakopoulos2024, Baake:2023zsq, Bakopoulos:2023sdm}. In these solutions, the hair explicitly manifests in both the spacetime metric and the scalar field configuration, distinguishing them from conventional black hole solutions in scalar-tensor theory.

To incorporate momentum dissipation into the model, we introduce two axionic fields that will explicitly break translational symmetry. Additionally, we include a Maxwell term to account for electromagnetic interactions. Consequently, the model we propose to study is described by the following four-dimensional action
\begin{eqnarray}
&&S=\int d^4x \sqrt{-g} \Bigg[
G_2+ G_4R + G_{4,X} \Big( (\Box\phi)^2 - \phi_{\mu\nu}\phi^{\mu\nu} \Big) \nonumber \\
&& 
+ F_4 \epsilon^{\mu\nu\rho\sigma} 
\epsilon^{\alpha\beta\gamma}_{\quad\,\sigma} 
\phi_{\mu}\phi_{\alpha}\phi_{\nu\beta}\phi_{\rho\gamma}
-\frac{1}{4}  F^{\mu \nu} F_{\mu \nu}\Bigg]\\
&&+\int d^4 x \sqrt{-g} \sum_{i=1}^2 \left( \frac{1}{2} \partial_\mu \psi_i \partial^\mu \psi_i \right)^k\nonumber
\label{actionbH}
\end{eqnarray}
where for simplicity we have defined $\phi_{\mu}=\partial_{\mu}\phi$
and $\phi_{\mu\nu}=\nabla_{\mu}\nabla_{\nu}\phi$, and where the
coupling functions $G_2$, $G_4$ and $F_4$ depend only on the kinetic
term $X=-\frac{1}{2}\phi_{\mu}\phi^{\mu}$ as 
{\small
\begin{eqnarray*}
&&G_2=-\frac{8 \eta}{3 \lambda^2} X^2, \qquad G_4=1-\frac{4 \eta}{3} X^2, \qquad F_4=\eta, \\
&&G_{4,X}:=\frac{d G_4}{dX}=-\frac{8 \eta}{3} X
\end{eqnarray*}}and where $\psi_i$  for $i=1, 2$ stand for the two axionic fields. We allow for the possibility of considering axionic fields in a k-essence form by introducing an exponent $k$. This generalization broadens the scope of our study, enabling us to investigate models where the dynamics of the axionic fields are governed by non-linear kinetic terms. Such k-essence frameworks provide additional flexibility in exploring the behavior of momentum dissipation mechanisms and their interaction with other components of the theory, such as the scalar field and the black hole hair. This inclusion also paves the way for examining richer phenomenology and potential observational signatures of these generalized models. For example, the case $k=2$ will be particular since the axionic part of the action will enjoy conformal symmetry, and this is achieved by $g_{\mu\nu}\to \Omega^2 g_{\mu\nu}$ and $\psi_i\to \psi_i$ for $i=1, 2$. 

In the following, we shall look for an ansatz of the form
\begin{eqnarray}
&&ds^2 = -h(r) d t^2 + \frac{d r^2}{f(r)} +r^2(dx_1^2+ d x_2^2), \\
&&\phi(t,r) = qt + \psi(r),\quad A_\mu d x^\mu=Q(r) d t,\quad \psi_i(x)=\omega\, x_i,\nonumber
\label{ansatztopo}
\end{eqnarray}
that aligns with the objectives we have described before. As shown below, the parameter $q$ will correspond to a primary hair. An ansatz of this kind for the scalar field was first considered in the context of stealth solutions for some specific sectors of Horndeski theory \cite{Babichev:2013cya}.

A solution within the ansatz \eqref{ansatztopo} to the field equations associated to the action \eqref{actionbH} is given by
\begin{eqnarray}
&&f(r)=h(r)=  -\frac{2\,M}{r}+\frac{2\,q^4\,\eta}{(r/\lambda)^2}+\frac{Q_e^2}{4\,r^2}-\frac{(\omega^2/2)^k \,}{(2k-3)\,r^{2(k-1)}},\nonumber\\
&& \left[\psi^{\prime}(r)\right]^2 = \frac{q^2}{h^2(r)}\left[1-\frac{h(r)}{(r / \lambda)^2}\right], \qquad Q(r)=-\frac{Q_e}{r}.
\label{Soluciones}
\end{eqnarray}
Various comments can be made concerning this axionic black hole solutions. First of all, in the absence of the electric charge and axion charge $Q_e=\omega=0$, the solution \eqref{Soluciones} corresponds to the hairy solution found in \cite{Bakopoulos:2023sdm} for a planar base manifold. Secondly, one could also consider the dyonic extension of the solution, although this is not necessary for our objective. Also, it can be observed that for a k-essence theory with $k=2$, corresponding to a conformally invariant axionic action, the term in the metric associated with the axionic charge decreases as $\sim \frac{1}{r^2}$, similar to the behavior of the Maxwell charge, which is also invariant in four dimensions. Finally, from the expression of the metric function $h(r)$, it appears that the case $k=\frac{3}{2}$ requires separate consideration. Specifically, one can see that for $k=\frac{3}{2}$, the metric solution exhibits a logarithmic dependence and is expressed as
\begin{eqnarray}
&&f(r)=h(r)=  -\frac{2\,M}{r}+\frac{2\,q^4\,\eta}{(r/\lambda)^2}+\frac{Q_e^2}{4\,r^2}+ \frac{\sqrt{2}\,\omega^3}{4}\frac{\ln(r)}{r},\nonumber\\
&& \left[\psi^{\prime}(r)\right]^2 = \frac{q^2}{h^2(r)}\left[1-\frac{h(r)}{(r / \lambda)^2}\right], \qquad Q(r)=-\frac{Q_e}{r}.
\label{Soluciones3/2}
\end{eqnarray}

\subsection{Holography DC Conductivity of the axionic solution}

In this subsection, we will focus on calculating the DC conductivity, following step by step the method outlined in \cite{Donos:2014cya} and arrive at the conclusion that the DC conductivity explicitly depends on the location of the horizon but not on the hair, even though the hair is explicitly present in the expression for the scalar field. The approach of Ref. \cite{Donos:2014cya} links the electrical conductivity of a boundary theory to data at the black hole horizon in the bulk spacetime. In doing so, we first consider small linearized perturbations of the background fields including some of the components of the metric, the Maxwell potential, and the two axionic fields as 
\begin{align}
      g_{tx_{i}}&=r^2\delta\,H_{tx_{i}}(r), &&g_{rx_{i}}=r^2 \delta\,H_{rx_{i}}(r),\\
       A_{x_i}&=\delta(-E_{x_{i}}\,t+a_{x_{i}}(r)),  &&\psi_{i}=\omega x_i+\delta\frac{\chi_{i}(r)}{\omega},\label{pertur}
\end{align}
where $a_{x_{i}}$, $H_{tx_{i}}$,  $H_{rx_{i}}$ and $\chi_{i}$ are independent fluctuation fields, and only depend on the radial coordinate. Secondly, we solve the linearized Maxwell equations, and this allows to define a conserved current $J_{{x_{i}}}$ in each spatial direction $i=1, 2$ as
\begin{equation}
\begin{aligned}
   &\partial_r\left(-h\,a'_{x_{i}}-Q'\,r^2\,H_{tx_{i}}\,\right)=0\Rightarrow\\
   &J_{{x_{i}}}:=-h\,a'_{x_{i}}-Q'\,r^2\,H_{tx_{i}}=-h\,a'_{x_{i}}+Q_{e}\,H_{tx_{i}}.\,
   \label{corriente}
\end{aligned}
\end{equation}
To ensure regular fluctuations, it is necessary to impose boundary conditions at the horizon and at infinity. It is convenient to use the Eddington-Finkelstein coordinates $(v,r)$ with $v=t+\int\frac{dr}{h(r)}$. In his coordinate system, the gauge field is then defined as
\begin{equation*}
    A_{x_i}=\delta\left(a_{x_i}-E\left(v-\int \frac{d r}{h(r)}\right)\right)d x_i,
\end{equation*}
and the fluctuated metric  becomes
\begin{equation*}
    \begin{aligned}
        ds^2 = -h d v^2 +  2\,d v d r +r^{2} &d x_1^2 + r^{2} d x_2^2 + \delta H_{t x_i} d v d x_i\\
        &- \frac{\delta H_{t x_i}}{h} d r d x_i + \delta H_{r x_i} d r dx_i.
    \end{aligned}
\end{equation*}
Upon substituting the previous conditions, the perturbed (metric) equation along the components $(r,x_{i})$, and denoted by  $\varepsilon_{rx_{i}}$, reads
\begin{equation}
\begin{aligned}
     \varepsilon_{rx_{i}}=\frac{Ex\,Q_e}{2\,r^2}&+H''_{tx_{i}}\left[\frac{ r^2\,h}{2}\right]+H'_{tx_{i}}\,2\,r\, h\\&-H_{tx_{i}}\,r^{2}\left[\varepsilon_{x_{i}x_{i}}+k\,\left(\frac{\omega ^2}{2\,r^{2}}\right)^k \right]=0.
\end{aligned}
\end{equation}
Here $\varepsilon_{x_{i}x_{i}}$, as before, stands for the metric equation along the components $(x_{i}, x_{i})$, and hence the condition $\varepsilon_{x_{i}x_{i}}=0$ holds. Now, as mentioned before, evaluating this expression at the location of the horizon $r_h$ of the metric function $h$, we abstract the differential equation to be solved and instead a obtain the expression for  
$H_{tx_{i}}$ which is given by 
\begin{equation}
    H_{tx_{i}}=\frac{2^{k-1}\,E_{x_{i}}\,Q_e}{k\,\omega^{2k}\,\,r^{2\left(2-k\right)}}\,\Big|_{r=r_{h}}.
\end{equation}
Finally, by substituting this expression into (\ref{corriente}), we find that the diagonal components of the conductivity matrix are given by
\begin{equation}
\sigma_{DC_{x_{i} x_{i}}}=\frac{\partial J_{x_{i}}}{\partial E_{x_{i}}}=1+\frac{2^{k-1}\,Q^2_e}{k\,\omega^{2k}\,\,r_h^{2(2-k)}}.\label{DC}
\end{equation}
Several observations can be made regarding this expression. The first, and by no means the least important, is that the expression does not explicitly depend on the hair $q$ even though the scalar field does. Secondly, in the case of the "standard" axionic coupling $k=1$, this expression matches exactly what is obtained in the context of a minimally coupled scalar field with axionic fields \cite{Donos:2014cya}. Additionally, for the conformal axionic coupling, i. e. $k=2$, the explicit dependence on the horizon vanishes, and the DC conductivity will depend only on the ratio between the electric charge and the axionic charge. As a final comment, it is worth noting that the value $k=3/2$ is not singular in the expression of the DC conductivity \eqref{DC}, unlike in the solution itself, see \eqref{Soluciones} and \eqref{Soluciones3/2}. Indeed, if the calculations for $k=3/2$ were carried out using the solution \eqref{Soluciones3/2}, one would obtain the expression of the DC conductivity \eqref{DC} with $k=3/2$.

The expression we obtained is, for the most part, nearly identical to the one derived in \cite{Donos:2013eha, Donos:2014cya} for a minimally coupled scalar field, possibly with an interaction potential. This suggests that, even when considering more general terms in the action, as we have done in this example, the calculation of the DC conductivity remains unaffected. This impression will be further reinforced in the more general example we propose to discuss in the following section. Building on these observations, in the next section, we will consider a more general beyond-Horndeski theory. Guided by the insights gained from our previous analysis, we will examine the implications of extending the framework while ensuring consistency with the key findings. This approach will allow us to confirm that the conclusions drawn earlier remain valid, even in a broader theoretical context.

\section{A more general set-up}
In order to extend our previous study to a more general model, we will consider the following Lagrangian
\begin{equation}
{\cal L}=\mathcal{L}_H+\mathcal{L}_{bH}-\frac{1}{4}F_{\mu\nu}F^{\mu\nu}+ \sum_{i=1}^2 \left( \frac{1}{2} \partial_\mu \psi_i \partial^\mu \psi_i \right)^k,\label{action}
\end{equation}
where we have defined 
\begin{equation}
\begin{aligned}
\mathcal{L}_{H}&=  G_2(\phi, X)-G_3(\phi, X) \square \phi+G_4(\phi, X) R\\
&+G_{4 X}\left[(\square \phi)^2-\phi^{\mu \nu} \phi_{\mu \nu}\right] +G_5(\phi, X) G^{\mu \nu} \phi_{\mu \nu}\\&-\frac{G_{5 X}}{6}\left[(\square \phi)^3-3 \square \phi \phi^{\mu \nu} \phi_{\mu \nu}+2 \phi_{\mu \nu} \phi^{\nu \lambda} \phi_\lambda^\mu\right],\\
\mathcal{L}_{bH}&= F_4(\phi,X) \epsilon^{\mu \nu \rho \sigma} \epsilon^{\alpha \beta \gamma}{ }_\sigma \phi_\mu \phi_\alpha \phi_{\nu \beta} \phi_{\rho \gamma},
\end{aligned}
\end{equation} where the coupling functions $G_i$ and $F_4$ can, a priori, be arbitrary functions of $\phi$ and the kinetic term $X$ as defined by $X=-g^{\mu \nu} \phi_\mu \phi_\nu / 2$. In this more general setup, we unfortunately do not have explicit solutions, but we will proceed as before by assuming that a solution exists within the previous ansatz \eqref{ansatztopo}, and where the Maxwell gauge field reads
\begin{equation}
 Q(r)=\int \frac{Q_e}{ r^{2} } \,\sqrt{\frac{h}{f}} \,dr.
\label{Sol2}
 \end{equation}
Furthermore, it is convenient to define the following function to considerably simplify the calculations, as was done in the case of the black hole solutions in \cite{Bakopoulos2024, Baake:2023zsq}
\begin{equation}
    \begin{aligned}
         W=G_4-&X\Bigg(4\,F_{4}\,X+2\,G_{4X}\\
         &-\sqrt{\frac{q^2-2\,h\,X}{h\,f}}\,\frac{f}{r}\,G_{5X}-G_{5\phi}\Bigg).\label{auxfuncion}
    \end{aligned}
\end{equation}
By performing the same calculation as before, the  conserved currents are given by 
\begin{equation}
\begin{aligned}
   &\partial_r\left(-\sqrt{h\,f}\,a'_{x_{i}}-Q'\,r^2\,H_{tx_{i}}\,\sqrt{\frac{f}{h}}\right)=0,\\
   &\Rightarrow \enspace J_{{x_{i}}}:=-\sqrt{h\,f}\,a'_{x_{i}}-Q'\,r^2\,H_{tx_{i}}\,\sqrt{\frac{f}{h}},\\
   &\quad\quad\quad\quad=-\sqrt{h\,f}\,a'_{x_{i}}-Q_{e}\,H_{tx_{i}}, \label{corriente2}
\end{aligned}
\end{equation}
The perturbed gauge field and metric are then given by 
\begin{equation*}
    A_{x_i}=\delta\left(a_{x_i}-E\left(v-\int \frac{d r}{\sqrt{h(r)\,f(r)}}\right)\right) d x_i,
\end{equation*}
and 
 \begin{equation}
    \begin{aligned}
        ds^2 =& -h d v^2 +  2\,\sqrt{\frac{h}{f}}\,d v d r + r^{2} d x_1^2 + r^{2} d x_2^2\\
        &+ \delta H_{t x_i} d v d x_i- \frac{\delta H_{t x_i}}{\sqrt{f\,h}} d r d x_i + \delta H_{r x_i} d r dx_i,
    \end{aligned} 
 \end{equation}
To ensure the regularity of the metric, one imposes that 
$$
H_{rx_{i}}=\frac{H_{t x_i}}{\sqrt{f\,h}},
$$
and, hence  the perturbed equation  $\varepsilon_{rx_{i}}$ is given by 
\begin{equation}
\begin{aligned}
    & \varepsilon_{rx_{i}}=-\frac{\,E_{x_i}\,Q_e}{2\,r^2}-H''_{tx_{i}}\left[\frac{ r^{2}\,h\,W}{2}\right]\\
     &-H'_{tx_{i}}\frac{ r^{2}}{2}\left[ \frac{h}{2}\left(\frac{f}{h}\right)'\,W+f\left(\,\frac{4\,W}{r}+W'\right) +\sqrt{\frac{f}{h}}\,\dot{W}\right]\\
     &+H_{tx_{i}}\,r^{2}\left[\varepsilon_{x_{i}x_{i}}+\,k\,\left(\frac{\omega ^2}{2\,r^{2}}\right)^k \right]=0.
\end{aligned}
\end{equation}
Here, as before the equation $\varepsilon_{x_{i}x_{i}}=0$ holds, and the prime (resp. dot) stands for derivative with respect to the radial (resp. to the $v$)  coordinate. The key novelty compared to the previous case lies in the evaluation of this expression at the horizon: the term proportional to $\dot{W}$ may no longer vanish if $W$ depends on the coordinate $v$ (which corresponds to $t$ in our original ansatz). This could happen if for example $W$ could depend explicitly on the scalar field. However, despite considering a general theory where the coupling functions could depend on both $\phi$ and $X$, solutions accommodating a scalar field of the form \eqref{ansatztopo} are generally associated with shift-invariant theories, where the coupling functions depend only on $X$, \cite{Bakopoulos2024, Baake:2023zsq}. Thus, without loss of generality, we can assume $\dot{W}=0$, and in this case, the evaluation at the horizon simplifies significantly yielding as before to  
\begin{equation}
    H_{tx_{i}}=-\frac{2^{k-1}\,E_{x_{i}}\,Q_e}{\,k\,\omega^{2k}\,\,r^{\left(4-2k\right)}}\,\Big|_{r=r_{h}}\label{Htx}
\end{equation}
Thus, by substituting this expression into (\ref{corriente2}), we find that the conductivity matrix is given as before 
\begin{equation}
\sigma_{DC_{x_{i} x_{i}}}=1+\frac{2^{k-1}\,Q^2_e}{\,k\,\omega^{2k}\,\,r_h^{\left(4-2k\right)}}.
\label{DC2}
\end{equation}
It is interesting to note that, even though we considered a more general action belonging to the beyond-Horndeski class, the perturbed equation $\varepsilon_{r x_i} = 0$ naturally includes a specific combination of the coupling functions, denoted as $W$ and defined in \eqref{auxfuncion}. Remarkably, when evaluated at the horizon, these terms vanish, leaving only contributions from the electric and axionic charges.  Consequently, we recover the same expression for the DC conductivity, namely an expression that does not depend in any way on the hair parameter $q$. This result confirms that the primary hair has no direct influence on the transport properties, with its role being entirely indirect through modifications to the horizon structure.

As we have just seen that the gravitational hair has no effect on conductivity, we now turn to the question of how the coupling of a scalar field to the Gauss-Bonnet invariant might influence conductivity. These models have recently attracted attention, with black hole solutions exhibiting properties inherent to the Gauss-Bonnet coupling being discovered and studied \cite{Sotiriou:2013qea, Sotiriou:2014pfa, Fernandes:2022zrq, Fernandes:2021dsb, Fernandes:2020rpa, Babichev:2022awg, Babichev:2023rhn, Babichev:2023dhs,Caceres:2023gfa}. These solutions can be incorporated within our action \eqref{action} in the absence of the beyond-Horndeski term $\mathcal{L}_{bH}=0$, considering a scalar field that depends only on the radial coordinate, that is $q=0$. The model we will study is described by the following Horndeski coupling functions 
\begin{equation}
\begin{aligned}
    G_2 &= -2 \Lambda - 2 \lambda e^{4 \phi} + 12 \beta e^{2 \phi} X + 8 \alpha X^2, \\
    G_3 &= 8 \alpha X, \\
    G_4 &= 1 - \beta e^{2 \phi} + 4 \alpha X, \\
    G_5 &= 4 \alpha \log X.
\end{aligned}
\end{equation}
and as discussed in \cite{Fernandes:2021dsb,Ayon-Beato:2023bzp}, this corresponds to the more general action leading to second-order field equations and whose scalar field equation is conformally invariant. In this setting, the metric solution $f$ solution  is given by
\begin{equation}
\resizebox{0.85\hsize}{!}{$
f(r) = \frac{r^2}{2 \alpha} \left( 1 \pm \sqrt{1 + 4 \alpha \left( \frac{2 M}{r^3} - \frac{Q_e^2}{2 r^4} + \frac{\Lambda}{3} + \frac{\left(\omega^2 / 2 r^2\right)^{k}}{(2k - 3)} \right)} \right),
$}
\end{equation}
together with the same axionic and gauge field as previously 
$$
A_{mu}dx^{\mu}=-\frac{Q_e}{r}dt,\qquad \psi_i(x)=\omega x_i.
$$
For the scalar field solution, we  have two possible branches. The first branch exists for $\lambda=0=\beta$ and given by
\begin{equation}
\phi(r)=\ln \left(\frac{c}{r}\right),\label{first branch}
\end{equation}
where c is integration constant. In this case, the auxiliary function $W$ takes the following form
\begin{equation}
W=\frac{r+2 \alpha f(r) \phi^{\prime}(r)\left(2+r \phi^{\prime}(r)\right)}{r},
\end{equation}
and the perturbed equation $\varepsilon_{rx_{i}}$ is then given by 
\begin{equation}
\begin{aligned}
    & \varepsilon_{rx_{i}}=-\frac{\,E_{x_i}\,Q_e}{2\,r^2}-H''_{tx_{i}}\left[\frac{ r^{2}\,f}{2}\left(1-\frac{2\,\alpha f}{r^2}\right)\right]\\
     &-H'_{tx_{i}}\frac{ r^{2}}{2}\left[f\left(\,\frac{4(r^2-\alpha f)-2\,r\,\alpha\,f'}{r^{3}}\right) \right]\\
     &+H_{tx_{i}}\,r^{2}\left[\varepsilon_{x_{i}x_{i}}+\,k\,\left(\frac{\omega ^2}{2\,r^{2}}\right)^k \right]=0\label{ferper}
\end{aligned}
\end{equation}
When evaluating this expression on the horizon, we obtain that the expression of $H_{tx_{i}}$ reduces to \eqref{Htx} and the conductivity matrix is given as before by \eqref{DC2}.

For the second branch exists provided that $\lambda=3\beta^2/ 4\,\alpha$, and in this case the scalar field reads
\begin{equation}
\phi(r)=\ln \left(\frac{ \sqrt{\frac{2\,\alpha}{\beta}}}{\left(c_{2} +\int \frac{1}{r \sqrt{\mathrm{f}(r)}} d r\right)}\right),\label{second branch}
\end{equation}
where $c_{2}$ is integration constant. In this case, the auxiliary function 
\begin{equation}
W=\frac{r-\beta \,e^{2\,\phi(r)}+2 \alpha f(r) \phi^{\prime}(r)\left(2+r \phi^{\prime}(r)\right)}{r}
\end{equation}
and as for the first branch,  we obtain the same expression as \eqref{DC2}. Hence, for this non-Noetherian conformal action \cite{Ayon-Beato:2023bzp}, the two branches of solutions yield the same expression of the conductivity which is clearly independent of the Gauss-Bonnet coupling $\alpha$.

\section{Conclusions and further prospects}

The paper explores the DC conductivity of black hole solutions in the context of beyond-Horndeski theories, with a particular focus on the role of primary hair. We have extended previous models by incorporating axionic fields to introduce momentum dissipation and a Maxwell term to account for electromagnetic effects. Despite the explicit inclusion of the scalar hair parameter $q$ in both the metric and the scalar field, our study demonstrates that the DC conductivity depends solely on the horizon location $r_h$ and not on the hair parameter itself.

The analysis has then be extended to a broader class of beyond-Horndeski theories, and the findings are consistent even when general coupling functions are considered. This result indicates that the scalar hair, while influencing the horizon structure, does not have a direct effect on the transport properties of the black hole solutions.

In accordance with current literature, we have demonstrated that the DC conductivity depends on the axionic charge (which is responsible for momentum dissipation) and the location of the horizon. An interesting observation is that when the action of the axionic fields is conformally invariant (the case $k=2$), this dependence on the horizon disappears. This raises an important question about the relationship between conformal invariance and the absence of horizon parameters in the DC conductivity. Understanding this relationship could provide deeper insights into how conformal symmetry affects transport properties, particularly in systems with momentum dissipation, and may help elucidate the role of axionic fields in holographic models. Specifically, further investigation is needed to explore why conformal invariance leads to a simplification where the horizon location no longer explicitly affects the conductivity, and what physical mechanisms are responsible for this phenomenon.

Future studies are suggested to examine the role of disformal transformations on axionic solutions, which could offer new insights into the effects of these transformations on the structure of black hole solutions and their corresponding transport properties. Indeed, we know that through a disformal transformation, solutions in the beyond-Horndeski class can be mapped to solutions belonging to the broader DHOST  class \cite{BenAchour:2016cay}. This raises the question of whether, in this more general framework, the primary hair could have a direct influence on transport properties. Exploring this possibility requires analyzing how the disformal transformation modifies the structure of the coupling functions and whether these modifications lead to explicit dependencies of transport coefficients on the hair parameter. It is not entirely clear whether disformal transformations, which are typically used to map solutions belonging to scalar-tensor theory classes, can be applied in a more general context. In our case, the models we are working with also involve an electromagnetic field and axionic fields. As a result, it is not obvious that disformal transformations can act within this broader setup. A first step in addressing this question would be to explore how solutions from scalar-tensor theories, which include both a Maxwell field and axionic fields, are mapped through a disformal transformation. This would help to clarify whether such transformations can be extended to these more complex models and provide a better understanding of their implications in modified gravity theories.

We have considered the possibility of a nonlinear action for the axions through the k-essence parameter, which impacts the DC conductivity. Similarly, we could have explored a nonlinear Maxwell Lagrangian as $\left(F_{\mu\nu}F_{\mu\nu}\right)^p$ \cite{Hassaine:2007py, Hassaine:2008pw}, where the parameter $p$ would also influence the DC conductivity calculation. This approach could highlight how modifications to the field's kinetic terms, whether via the axionic or electromagnetic sector, can affect transport properties in holographic models. In the same register, it would be interesting to consider more general nonlinear electrodynamics, for which black hole solutions have recently been studied, see e. g. \cite{Ayon-Beato:2024vph,Barrientos:2024umq}. It would be worthwhile to investigate whether their extensions remain viable in the presence of axions.

The initial objective of this work was to determine the extent to which the primary hair of axionic black hole solutions could influence holographic properties through the calculation of transport coefficients. Through our analysis, we have become convinced that no matter how general the theories accommodating these hairy solutions may be, the DC conductivity remains insensitive to the hair parameter. Nevertheless, it would be interesting to have a formal proof on our intuition, and we leave it for a further work.

\begin{acknowledgments}
  This work has been partially funded by  ANID Grant No.~21231297 
\end{acknowledgments}

\bibliographystyle{apsrev4-2}
\bibliography{References}

\end{document}